\documentclass [aps,prb,twocolumn,amssymb,showpacs,superscriptaddress,reprint,citeautoscript]{revtex4-1}

\usepackage{amsfonts}
\usepackage{amsmath}
\usepackage{amssymb}
\usepackage[pdftex]{graphicx}
\usepackage{bm}

\begin{document}



\title{ Upper critical field of high quality single crystals of KFe$_2$As$_2$ }


\author{Yong Liu}
\email{yliu@ameslab.gov}
\affiliation{Ames Laboratory, Ames, Iowa 50011, USA}

\author{M.~A.~Tanatar}
\email{tanatar@ameslab.gov}
\affiliation{Ames Laboratory, Ames, Iowa 50011, USA}
\affiliation{Department of Physics and Astronomy, Iowa State University, Ames, Iowa 50011, USA }

\author{V.~G.~Kogan}
\email{kogan@ameslab.gov}
\affiliation{Ames Laboratory, Ames, Iowa 50011, USA}

\author{Hyunsoo Kim}
\email{hyunsoo@iastate.edu}
\affiliation{Ames Laboratory, Ames, Iowa 50011, USA}
\affiliation{Department of Physics and Astronomy, Iowa State University, Ames, Iowa 50011, USA }

\author{T.~A.~Lograsso}
\email{lograsso@ameslab.gov}
\affiliation{Ames Laboratory, Ames, Iowa 50011, USA}

\author{R.~Prozorov}
\email[Corresponding author: ]{prozorov@ameslab.gov}
\affiliation{Ames Laboratory, Ames, Iowa 50011, USA}
\affiliation{Department of Physics and Astronomy, Iowa State University, Ames, Iowa 50011, USA }

\date{4 April 2013}


\begin{abstract}

Measurements of temperature-dependent in-plane resistivity , $\rho(T)$, were used to determine the upper critical field and its anisotropy in high quality single crystals of stoichiometric iron arsenide superconductor KFe$_2$As$_2$. The crystals were characterized by residual resistivity ratio, $\rho(300K)/\rho(0)$ up to 3000 and resistive transition midpoint temperature, $T_c$=3.8~K, significantly higher than in previous studies on the same material. We find increased $H_{c2}(T)$ for both directions of the magnetic field, which scale with the increased $T_c$. This unusual linear $H_{c2}(T_c)$ scaling is not expected for orbital limiting mechanism of the upper critical field in clean materials.

\end{abstract}





\maketitle



\section{Introduction}

Among several families of iron arsenides showing superconductivity at temperatures up to 56~K \cite{Paglione,CanfieldBudko,Johnston,Stewart}, very few compounds are stoichiometric. Due to the lack of substitution disorder, these compounds reveal the properties of true clean materials and are characterized by significantly enhanced residual resistivity ratios, $rrr \equiv \rho(300K)/\rho(T_c)$ up to $\sim$80 in LiFeAs ($T_c \approx$18~K) \cite{LiFeAs}, $rrr \sim$10 in environmentally-doped NaFeAs ($T_c \approx 25~K$) \cite{TanatarNaFeAs} and up to 1500 in KFe$_2$As$_2$ (K122 in the following) \cite{ReidK}. The superconducting $T_c$ and $rrr$ of the latter material strongly vary\cite{FukazawaNMRHC,Dong,HashimotoK} depending upon sample quality and preparation technique, and are very sensitive to doping with Co \cite{ChenCodoping} and Na \cite{ChenglinNa}, suggestive of unconventional superconductivity. 
Indeed, all studies of the  superconducting gap structure in K122 agree on the existence of line nodes \cite{FukazawaNMRHC,Dong,ChenCodoping,ChenglinNa,HashimotoK,ReidK,Kawano,ARPES}, however, their location on the multi-band Fermi surface, symmetry-imposed vs. accidental character and relation to $S \pm$ or $d-$wave symmetry \cite{Thomale,Korshunov} are highly debated. 

In this study we use high sensitivity of the superconducting transition temperature of KFe$_2$As$_2$ to residual impurities to obtain an insight into another unique feature of iron arsenide superconductors,- unusual temperature dependence of the upper critical field. Terashima \textit{et al.} \cite{TerashimaHc2} reported anisotropic $H_{c2}$ of the K122 crystals with $T_c$=2.8~K, which revealed very different temperature dependence for magnetic fields parallel and perpendicular to the $c$-axis of the tetragonal crystal, with virtually $T$-linear dependence for $H \parallel c$. This dependence is different from the expectations of theories for orbital Werthamer, Helfand and Hohenberg (WHH) \cite{WHH} and paramagnetic \cite{CC} mechanisms of $H_{c2}$, both predicting saturation of $H_{c2}(T)$ on $T \to 0$. It is also strongly different from saturating $H_{c2}$ found in LiFeAs \cite{KyuilHc2LiFeAs,LiFeAsTerashima,Balakirev}. Recently we found similar $T$-linear dependence in the optimally-doped SrFe$_2$(As$_{1-x}$P$_x$)$_2$, $x$=0.35 \cite{SteveHc2}. Because of the nodal superconducting gap of this compound \cite{SrPTDR}, we speculated possible link of nodal superconducting gap and the $T$-linear dependence of $H_{c2}$. On the other hand, $T$-linear dependence of $H_{c2,c}(T)$ observed in dirty iron-pnictides \cite{Gurevich} and doped MgB$_2$ \cite{GurevichMgB2} was explained in the orbital-limiting model for two-band superconductivity in the dirty limit, as a cross-over regime between usual WHH saturating and upward curving dependences.

In this article, we report synthesis of single crystals of KFe$_2$As$_2$  with $rrr$ up to about 3000 and study their anisotropic upper critical field. 
We find higher $H_{c2}$ for both directions of magnetic field than found by Terashima \textit{et al.} \cite{TerashimaHc2}. Interestingly, the two data sets for this material, which in both high and low quality samples is in the clean limit, can be matched by a factor corresponding to $T_c$ ratio. 
As we show, this unusual linear dependence between $H_{c2}(0)$ and $T_c$ is not expected in any theory for clean superconductors for orbital limiting mechanism. We discuss possible important parameters for its explanation.

\section{Experimental}

\begin{figure}[tb]%
\centering
\includegraphics[width=8cm]{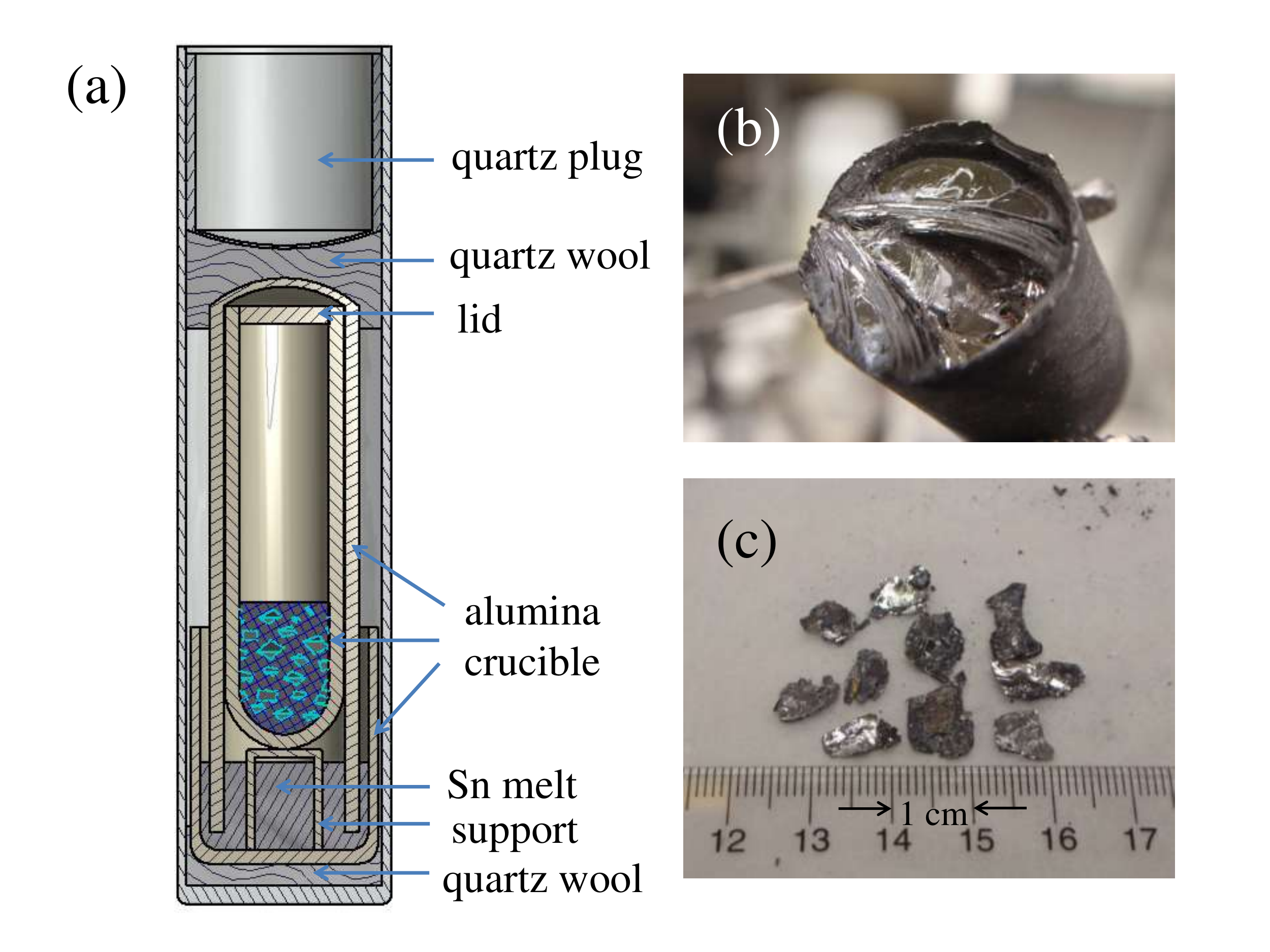}%
\caption{(Color online) (a) Crystal growth setup with liquid Sn-melt sealing technique. (b) Top view of KFe$_2$As$_2$ ingot of 15 mm diameter, revealing easy to distinguish the shiny pieces and lamellar structure of crystals. (c) Single crystals of KFe$_2$As$_2$, cleaved out of the ingot, with sizes up to 10$\times$5$\times$0.2 mm$^3$.
}%
\label{crystals}%
\end{figure}

Single crystals of KFe$_2$As$_2$ were grown using the KAs flux method \cite{crystals}. It is difficult to grow KFe$_2$As$_2$ by sealing the chemicals in a quartz ampoule because of the strong reaction between the potassium vapor and the silica tube, leading to a serious corrosion. As a way to avoid this problem, Kihou {\it et al.} \cite{crystals} suggested use of stainless steel containers. Alternatively, we developed a sealing technique with liquid tin melt to suppress the evaporation of potassium and arsenic chemicals \cite{YLiu}. K ingot, As lump, and Fe powder were mixed in atomic ratio of K:Fe:As=5:2:6, and loaded into an alumina crucible. This crucible was covered by a bigger crucible and then mounted in a third alumina crucible with the Sn chunk spread on the bottom, as shown in Fig. \ref{crystals}(a). The effectiveness of Sn-melt sealing technique is guaranteed by the fact that Sn melts at low temperature 232$^{\circ}$C, but boils at high temperature 2602$^{\circ}$C. On the other hand, Sn melt acts as a buffer, which can also dissolve part of potassium  and arsenic vapors. By sealing the apparatus shown in Fig.~\ref{crystals}(a) in a bigger quartz tube, we could grow large crystals, as shown in Figs.\ref{crystals} (b) and (c), using cooling at a rate of 4$^{\circ}$C/h from 920$^{\circ}$C to 820$^{\circ}$C and at 1 $^{\circ}$C /h from 820$^{\circ}$C to 620 $^{\circ}$C.

\begin{figure}[tb]%
\centering
\includegraphics[width=8cm]{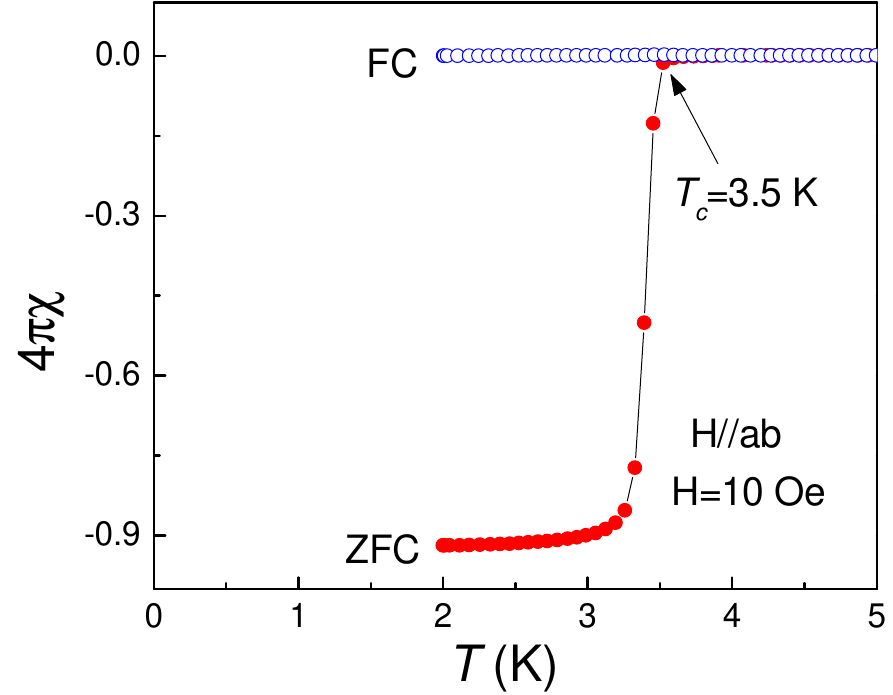}%
\caption{(Color online) (a) DC SQUID magnetization measurements of the KFe$_2$As$_2$ crystals in zero field cooled (ZFC) and field cooled (FC) protocols in a magnetic field of 10 Oe applied parallel to the $ab$-plane. 
}%
\label{MT}%
\end{figure}

\begin{figure}[tb]%
\centering
\includegraphics[width=8cm]{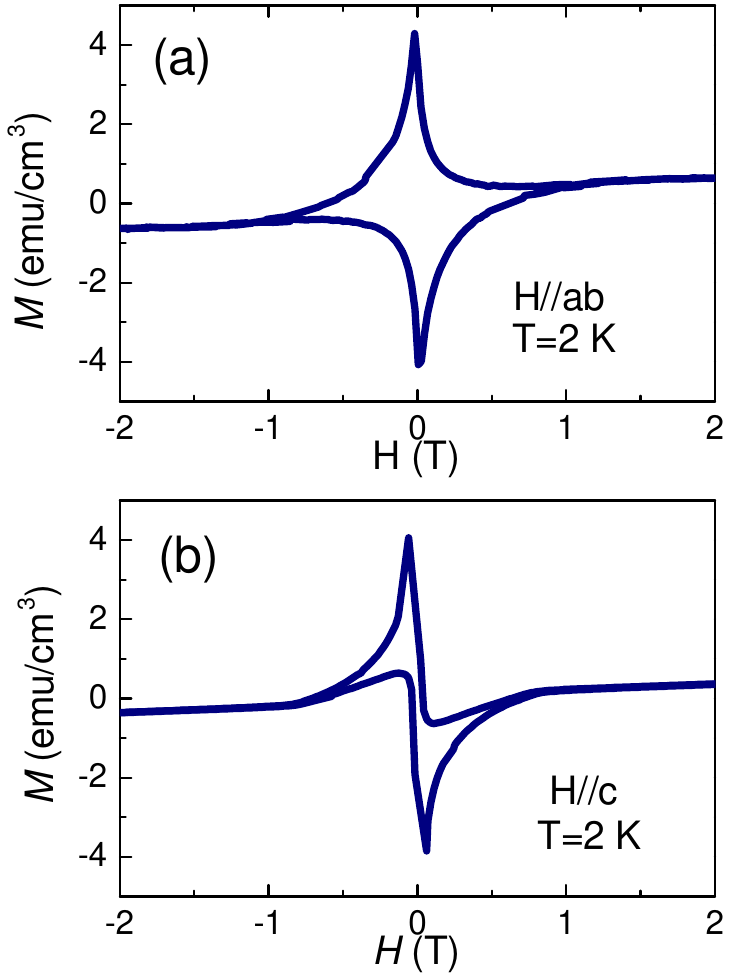}%
\caption{(Color online) DC SQUID measurements of magnetization loops at 2~K, the base temperature of our SQUID apparatus, in magnetic fields applied parallel to the $ab$ plane (a) and parallel to the tetragonal $c$-axis (b). 
}%
\label{MH}%
\end{figure}

Crystals with size up to $10 \times 5 \times 0.2$~mm$^3$ were extracted from the melt and frequently had leftover KAs flux on the surface. Its presence causes rapid sample degradation in air. The crystals were characterized by MPMS DC SQUID magnetization measurements, as shown in Figs.~\ref{MT} and \ref{MH}.

Figure \ref{MT} shows temperature-dependent magnetization measured after cooling in zero field, applying a 10 Oe field at base temperature and making measurements on warming above the superconducting transition (zero - field cooling, ZFC) and after cooling in the same magnetic field and measurements on warming (field cooling, FC). Sharp superconducting transition with the width of about 0.3 K in zero-field cooling measurements shows high quality of the single crystals. The field cooling results are similar to other iron-based superconductors showing the absence of the Meissner expulsion in FC measurements, implying an anomalous Meissner effect \cite{Prozorov2010b}.

Figure~\ref{MH} shows magnetization loops measured at $T=$2~K along the $ab-$plane (panel a) and along the $c-$axis (panel b). A clear contribution of asymmetric reversible magnetization and a sharp break at the $H_{c2}$ in the latter measurements indicate low pinning; hence high sample quality consistent with a large $rrr_a$ = 2500-3000. The pinning is notably larger along the conducting planes, probably indicating the intrinsic pinning contribution on the layered structure, also consistent with a much lower $rrr_c$.

During sample preparation for resistivity measurements, we first cleaved slabs from the inner parts of single crystals. These crystal slabs with two cleaved mirror-like surfaces turned out relatively stable. The slabs were further cleaved into bars with typical dimensions of (1- 2)$\times$0.5$\times$(0.02-0.1) mm$^3$ and long axes parallel to $a-$ crystallographic direction. All sample dimensions were measured with an accuracy of about 10\%. Contacts for four-probe resistivity measurements were made by soldering 50 $\mu$m silver wires with ultrapure Sn solder, as described in Ref.~\onlinecite{SUST}. This technique produced contact resistance typically in the 10 $\mu \Omega$ range. Resistivity measurements were made in the {\it Quantum Design} PPMS system. 

The resistivity value at room temperature for our samples was about 300 $\mu \Omega$cm, similar to the values found in previous studies for KFe$_2$As$_2$ \cite{crystals,ReidK}, and the value in slightly doped BaK122 \cite{Wenrhoa}. This slight variation of $\rho(300K)$ in the (Ba,K)Fe$_2$As$_2$ system with doping is distinctly different from a rapid decrease of $\rho(300K)$ with $x$ in Ba(Fe$_{1-x}$Co$_x$)$_2$As$_2$ \cite{pseudogap}. After removing parts of the crystal exposed to the KAs flux, the samples became relatively stable and their resistivity did not change for a period of a week or so. 
We selectively measured inter-plane resistivity of some samples using the two-probe technique. The details of the measurement procedure for in-plane and inter-plane, $\rho_c (T)$, resistivity measurements can be found in Refs.~\onlinecite{anisotropy,anisotropypure,pseudogap}. 

Measurements of the upper critical field were made by gluing the sample for $\rho_a$ measurements with Apiezon N grease to a top or side surfaces of a G10 plastic cube, enabling precise orientation of the sample plane parallel and perpendicular to the magnetic field with an accuracy of about 1$^{\circ}$. Considering relatively flat dependence of $H_{c2}$ on the field inclination angle $\theta$ in this small angular range \cite{TerashimaHc2}, this alignment procedure is sufficiently precise. Measurements were completed in two transverse field vs. current configurations $J \parallel a$, $H \parallel c$, and $J \parallel a$, $H \parallel b$.

\section{Results}

\subsection{Residual resistivity ratio}

\begin{figure}[tb]%
\centering
\includegraphics[width=8cm]{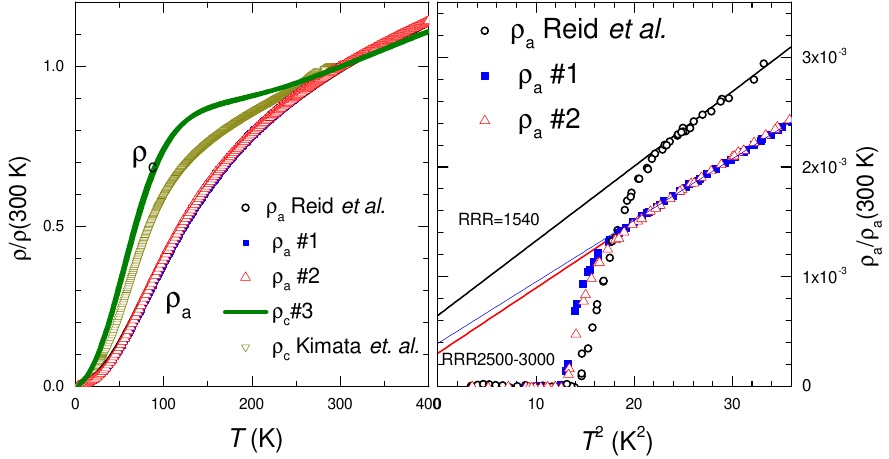}%
\caption{(Color online) Temperature-dependent electrical resistivity of KFe$_2$As$_2$ for current direction along the plane in samples  1 and 2 (this study) in comparison with published data with highest residual resistivity ratio by Reid. \textit{et al.} \cite{ReidK} Left panel shows data over a broad temperature range, right panel shows the same data plotted vs $T^2$, allowing for linear fits through data above the onset of the resistive transition and extrapolation of the $\rho(T)$ to $T=0$ to determine $\rho (0)$. Samples  1 and 2 show lower residual resistivity values at $T_c$ and in $\rho(0)$ extrapolation than the best samples reported so far by Reid \textit{et al.},  giving $\rho(300~\mathnormal{K})/\rho(0)$ in 2500 to 3000 range, significantly higher than in all previous reports \cite{ReidK,TerashimaHc2,crystals,Dong,HashimotoK}. For reference in left panel we show $\rho_c(T)$ as measured in sample 3 in this study and reported by Kimata \textit{et al.}, Ref.~[\onlinecite{Kimata}].
}%
\label{resistivity}%
\end{figure}

The left panel of Fig.~\ref{resistivity} shows temperature-dependent resistivity over a broad temperature range. The right panel shows a zoom of the low-temperature portion, with the data plotted vs. $T^2$, in the vicinity of the superconducting transition. For reference we show the $\rho_a(T)$ data for the samples with the highest $rrr$ among previously published data,\cite{ReidK} and inter-plane resistivity measured in the sample  3 of our batch in comparison with measurements by Kimata {\it et al.} \cite{Kimata} First we notice that the samples grown in different labs show identical $\rho_a(T)$, except for the variation of residual resistivity $\rho_a(0)$. This fact clearly shows that the difference is caused by very low density of residual impurities/defects, determined by the growth technique, but not variation of sample composition/stochiometry. The defects determining residual resistivity reflect uncontrolled sample chemical contamination during growth and density of non-equilibrium vacancy-interstitial defects at the growth temperature. The density of these defects can be estimated from comparison of the mean free path in our samples ($>$1000 nm), see below, and lattice constants ($\sim$1 nm), as $\sim 10^{17}$cm$^{-3}$. This is negligible to produce any doping in a good metals like KFe$_2$As$_2$ with carrier density higher than 10$^{21}$cm$^{-3}$.

While showing complicated temperature dependence over a broad temperature range, with notable crossover at around 200~K, the resistivity at the lowest temperatures follows simple close to $T^2$ dependence, \cite{ReidK} as expected in Landau Fermi liquid theory. This temperature dependence is most easily seen when plotting $\rho$ vs $T^2$ as shown in the right panel, providing a linear plot. In Fermi liquid theory, the slope of the curves, $A$, is proportional to the square of effective mass, $A \sim m*^2$. As can be seen from Fig.~\ref{resistivity}, this slope remains the same, within error bars, for both high and low quality samples, showing directly that the difference in sample quality is not related to variation of effective mass and hence band structure. This simple linear dependence of $\rho$ vs $T^2$ enables easy extrapolation of the $\rho (T)$ from $T_c$ to $T=0$. As can be seen from the right panel in Fig.~\ref{resistivity}, samples  1 and  2 show nearly indistinguishable temperature-dependent resistivity with both $\rho(T_c)$ and extrapolated $\rho (0)$ significantly lower than in samples by Reid {\it et al.}, leading to a residual resistivity ratio $\rho(300K)/\rho(0)$ in the 2500 to 3000 range.

\subsection{Upper critical field}

\begin{figure}[tb]%
\centering
\includegraphics[width=8cm]{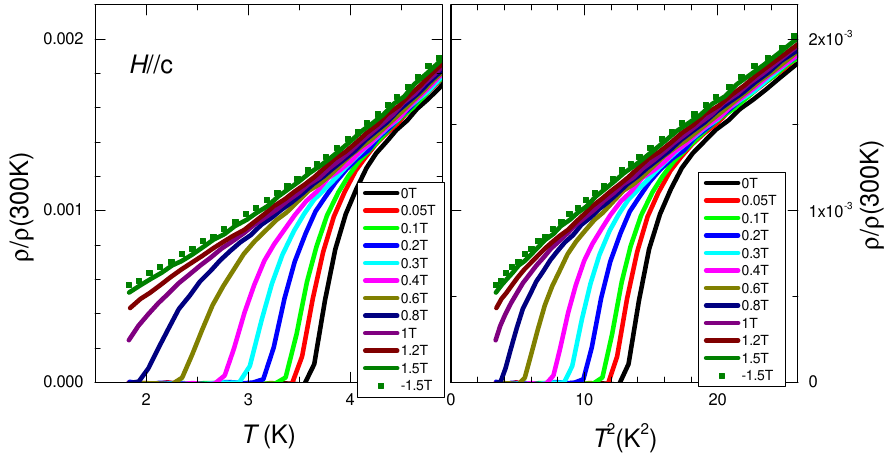}%
\caption{(Color online) 
Temperature-dependent in-plane electrical resistivity of KFe$_2$As$_2$ sample  1 in magnetic fields aligned parallel to the $c-$axis. Left panel shows $\rho_a/\rho_a(300K)$ plotted vs. $T$, right panel shows the same data plotted vs $T^2$. Field values increase from 0 to 1.5~T. Lines and symbols at the highest field of 1.5~T show resistivity for two reversed directions of magnetic field, revealing no significant contribution of the Hall voltage to resistivity measurements. Note, at 1.5~T and 1.8~K, the base temperature of our apparatus, the $\rho_a(T)$ does not show any signature of saturation. The actually measured $rrr$ is about 2000. 
}
\label{pp}%
\end{figure}

\begin{figure}[tb]%
\centering
\includegraphics[width=8cm]{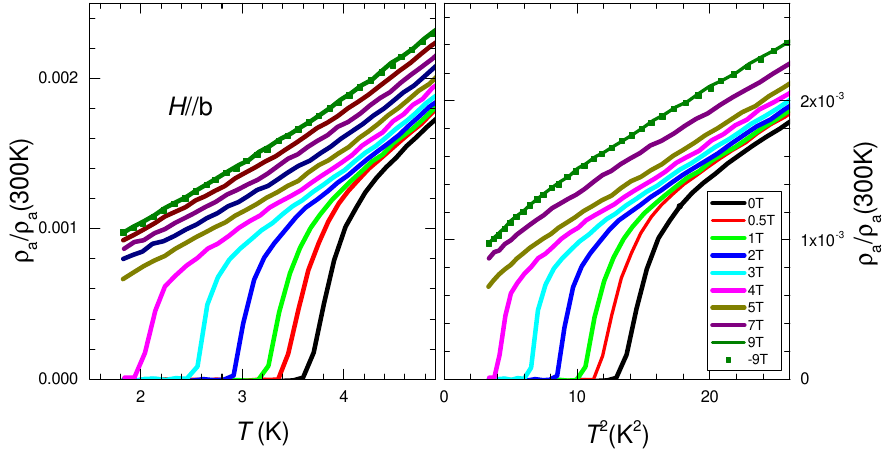}%
\caption{(Color online) 
Temperature-dependent in-plane electrical resistivity of KFe$_2$As$_2$ sample  1 in magnetic fields aligned in-plane transverse to the current parallel to the $b$-axis. The left panel shows $\rho_a/\rho_a(300K)$ plotted vs. $T$. The right panel shows same data plotted vs $T^2$. Field values increase from 0 to 9~T. Lines and symbols at the highest field of 9~T show resistivity for the two reversed directions of magnetic field, revealing negligible contribution of the Hall voltage to the resistivity measurements. Notable magnetoresistance is observed in this configuration, which leads to clear downward deviations from perfect $T^2$ dependence observed at low fields. 
}
\label{ll}%
\end{figure}

Figure~\ref{pp} shows temperature-dependent in-plane electrical resistivity of KFe$_2$As$_2$ sample  1 in magnetic fields aligned parallel to tetragonal $c$-axis. Left panel shows $\rho_a/\rho_a(300K)$ plotted vs. $T$, right panel shows the same data plotted vs $T^2$. Magnetic field values increase from 0 to 1.5~T, sufficient to completely suppress superconductivity at $T>$1.8~K, the base temperature of our apparatus. To check if resistivity measurements in these very low resistivity samples with $\rho(0)$ in 100 to 200 $n\Omega.$cm range contain contributions from Hall voltage, we reversed the direction of the magnetic field at the highest field of 1.5~T, with the data shown with line and symbols. It is clear the Hall contribution is insignificant in our measurements. 

The high purity of our samples can be directly seen from the fact that even at 1.8~K, $\rho_a(T)$ does not show any sign of saturation, and actually measured resistivity $\rho(1.8K,H=1.5T)$ gives $rrr \approx$ 2000. A deviation of the $\rho(T^2)$ plot from linear can be noticed in the right panel of Fig.~\ref{pp} for non-zero magnetic fields. While zero-field data follow $T^2$ temperature-dependence above $T_c$, a crossover with downward curvature at around 2.5~K is seen in $\rho(T)$ at 1.5~T. This crossover is more evident in measurements in $H \parallel b$ around 3~K, see Fig.~\ref{ll}. This crossover feature was reported as an indication of non-Fermi-liquid dependence in initial measurements by Dong {\it et al.}, \cite{Dong} however, it was suggested to be superconducting in origin by later experiments \cite{Terashimacomment}. 

Measurements with magnetic field reversal at 9~T, Fig.~\ref{ll}, show that contribution of Hall voltage in $H \parallel b$ configuration is significantly smaller than at 1.6~T in the $H \parallel c$ configuration, Fig.~\ref{pp}, and can be safely neglected.

\section{Discussion} 


\begin{figure}[tb]%
\centering
\includegraphics[width=8cm]{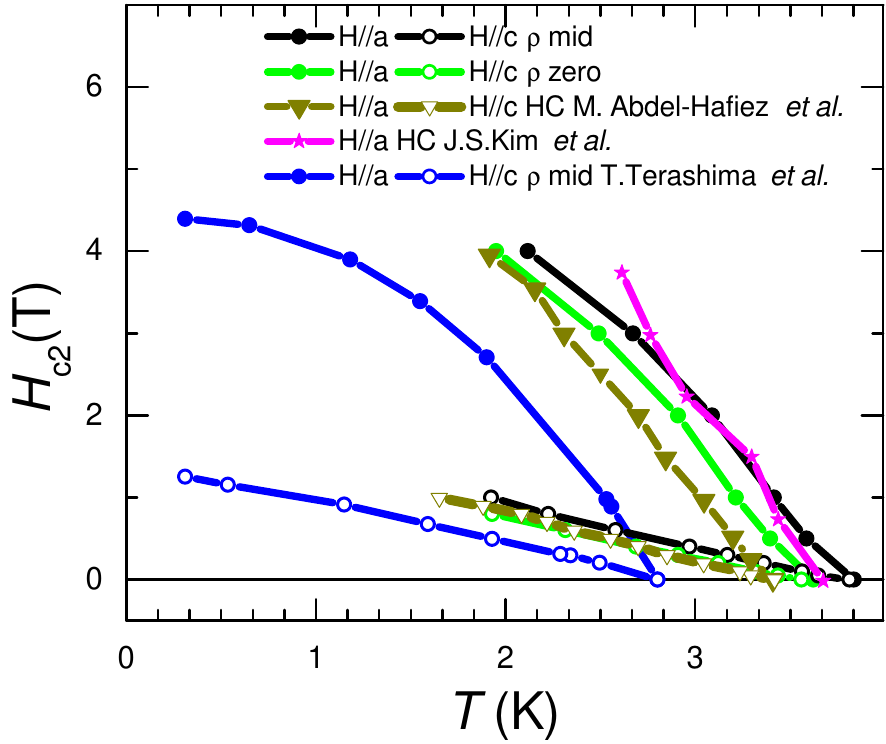}%
\caption{(Color online) The upper critical field as determined from the mid-point (black curves) and offset (green curves) of resistive transition in the KFe$_2$As$_2$ sample 1. Open symbols show data for $H \parallel c$, solid symbols for $H \parallel b$. Blue curves show data determined from resistive transition midpoint in Ref.~\onlinecite{TerashimaHc2}, from measurements on lower quality samples. Green curves with down triangles show data determined from specific heat measurements by Abdel-Hafiz {\it et al.} \cite{Hc2Buechner}, magenta curve is from specific heat measurements in the magnetic field parallel to the plane by Kim {\it et al.} \cite{StewartHc2K}  }%
\label{HTdiagram}%
\end{figure}

\begin{figure}[tb]%
\centering
\includegraphics[width=8cm]{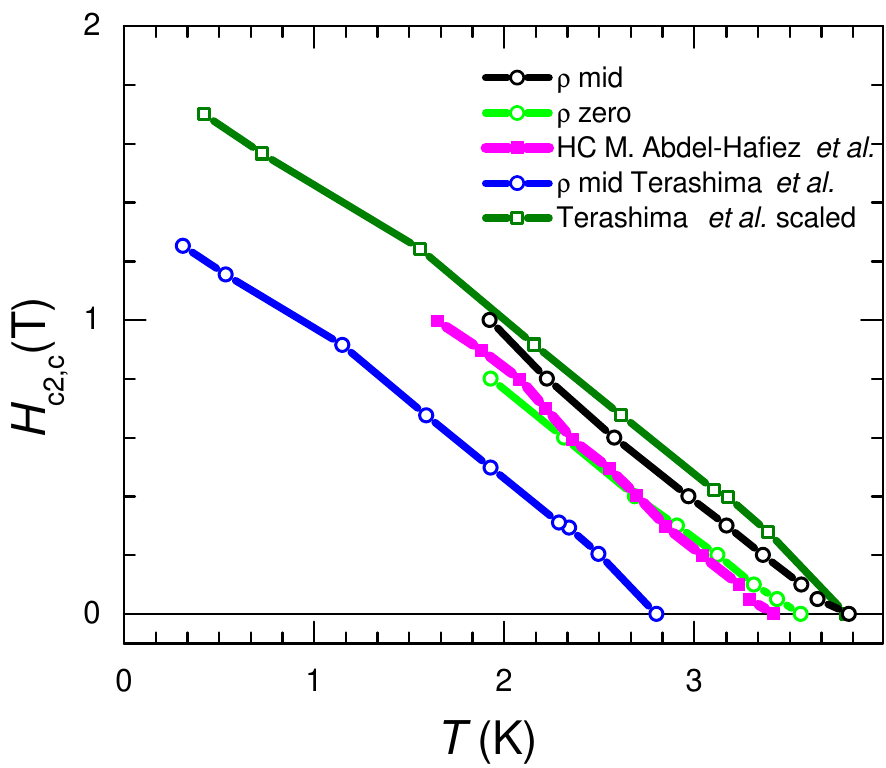}%
\caption{(Color online) The upper critical field in configuration $H \parallel c$ as determined from mid-point (black curves) and offset-point (green curves) of resistive transition in KFe$_2$As$_2$ sample  1. Blue curves show data determined from resistive transition midpoint by Terashima {\it et al.}\cite{TerashimaHc2} from measurements on lower quality samples. Cyan curves show the data from Ref.~\onlinecite{TerashimaHc2} with $H_{c2,a}$, $H_{c2,c}$, and $T_c$ multiplied by a constant factor to match $T_c$ of high quality samples.    
For reference, we show data determined from specific heat measurements by Abdel-Hafiz {\it et al.}\cite{Hc2Buechner} Note, irrespective of the measurements type or criteria used, the slope of the curves does not depend on sample $T_c$.  }%
\label{comparison}%
\end{figure}

In Fig.~\ref{HTdiagram}, we show the magnetic field-temperature phase diagram as determined in our measurements on high quality single crystals, in comparison with measurements on low quality samples in  Ref.~[\onlinecite{TerashimaHc2}]. The data were determined using a commonly accepted resistive transition midpoint $\rho _{mid}$ and offset criteria. The former is identical to the criteria used by Terashima {\it et al.} \cite{TerashimaHc2} 

Using midpoint criterion, we obtain the superconducting $T_c$=3.8~K in zero field, as compared to 2.8~K for samples in Terashima {\it et al.} study. \cite{TerashimaHc2} This leads to notably higher $H_{c2}$ for both principal directions of the magnetic field. To obtain further insight into the behavior of $H_{c2}$ of KFe$_2$As$_2$, we compare our measurements on highest quality samples with two recent heat capacity studies undertaken on high quality samples ($rrr \sim$600) \cite{Hc2Buechner,StewartHc2K}. These measurements show agreement of two measurement types, despite very different criteria. In view of possible misalignment of the field parallel to the plane, we will focus below on $H \parallel c$, shown in Fig.~\ref{comparison}.

Interestingly, the slope of the $H_{c2}(T)$ curve at $T_c$ turns out to be the same in all measurements. Moreover, by multiplying both $T_c$ and $H_{c2}$ obtained in Terashima {\it et al.} \cite{TerashimaHc2}  study by a factor of $T_c$ ratio, $\sim$1.36, we obtain a good match of the data  for $H \parallel c$ and reasonable matching for $H \parallel a$. 

For our understanding, it is important that both low quality samples with $rrr=$80 and extreme quality samples, $rrr \sim$2000-3000, are in the clean limit. Indeed, the mean free path of the dirty samples can be estimated as 100 nm, assuming $v_f=2\times 10^5$ m/s, it is significantly higher in high quality samples. Simultaneously, the coherence length can be estimated from zero-temperature value of $H_{c2}$ using the relation $H_{c2}(0)=\phi_0/(2 \pi \xi^2_0)$, where $\phi_0 = 2.07 \times 10^{-7}$~G cm$^2$ is magnetic flux quantum, $\xi _0$ is the coherence length at $T=0$. For $H_{c2}(0)=$1.5~T, this gives $\xi_0=$20 nm, significantly smaller than the mean free path, so all samples are in the clean regime.

In clean isotropic superconductors, the zero-$T$ upper critical field and its slope at $T_c$ scale as \cite{WHH}

\begin{align}
H_{c2}(0) \propto \frac{\phi_0T_c^2}{\hbar^2 v_F^2} \,,\qquad 
\frac{dH_{c2}}{dT}\Big|_{T_c} \propto \frac{\phi_0T_c }{\hbar^2 v_F^2}\,.
\label{scaling}
\end{align}

These scalings hold -- in clean case -- also for anisotropic order parameters on anisotropic Fermi surfaces.\cite{KoganProzorov} One can write Eq.\,(\ref{scaling}) as 
 
\begin{align}
H_{c2}(0) \propto \frac{ T_c^2}{E_F} \,,\qquad \frac{dH_{c2}}{dT}\Big|_{T_c} \propto \frac{ T_c }{E_F }\,.
\label{scaling}
\end{align}

 Our results, $H_{c2}(0) \propto T_c$ and $H_{c2}^\prime(T_c) = $ const, suggest  a curious possibility that $T_c$ in our set of clean samples of KFe$_2$As$_2$ is proportional to the Fermi energy $E_F$. We are not aware of a theoretical argument in favor of this possibility.


\section{Conclusions}

Measurements of the in-plane electrical resistivity as a function of the magnetic field applied parallel and perpendicular to the tetragonal $c$-axis of the crystal allow us to extrapolate residual resistivity of the samples in the zero field in 100 to 200 $n \Omega.$cm range and residual resistivity ratio in the 2500 to 3000 range. These high values are in reasonable agreement with resistivity measurements in the normal state achieved by application of the magnetic field 1.5~T $H\parallel c$ at base temperature of 1.8~K, $rrr \sim $2000. 

The upper critical fields in our samples with $T_c$=3.8~K significantly increased, compared to those for samples with $T_c$=2.8~K, but $H_{c2}$ for two sets of samples can be matched well by a simple scaling of $T_c$. This unusual linear relation is not expected for the orbital limiting mechanism of the upper critical field.

\section{Acknowledgments}
The work at Ames was supported by the U.S. Department of Energy, Office of Basic Energy Sciences, Division of Materials Sciences and Engineering under contract No. DE-AC02-07CH11358. 


\end{document}